\documentstyle[11pt,newpasp,twoside,epsf]{article}
\def\edcomment#1{\iffalse\marginpar{\raggedright\sl#1\/}\else\relax\fi}
\reversemarginpar

\begin{document}
\title{An Automated Search for Extrasolar Planet Transits}
  \author{M. G. Hidas, J. K. Webb, M. C. B. Ashley, C. H. Lineweaver}
\affil{University of New South Wales, Sydney 2052 NSW, Australia}
\author{J. Anderson}
\affil{University of California, Berkeley, USA}
\author{M. Irwin}
\affil{Institute of Astronomy, Cambridge, UK}

\begin{abstract}
We are setting up a new search for transiting extra-solar planets using the 0.5m Automated Patrol Telescope at Siding Spring Observatory, Australia. We will begin regular observations in September 2002. We expect to find $\sim7$ new planets per year.
\end{abstract}

\section{Introduction}

An increasing number of teams are searching for extra-solar planets using the transit method (see Horne 2002 for a review). Although the probability of observing a transit for any given star is small, using a wide-field telescope a large number of stars can be monitored, potentially yielding a higher detection rate than the radial velocity surveys (e.g. Marcy et al. 2002; Mayor et al. 2002). Furthermore, for a transiting planet orbiting a sufficiently bright star, the planet's size, as well as its actual mass and orbital characteristics (from follow-up spectroscopy) can be determined, constraining models of its structure and formation.

\section{The Automated Patrol Telescope}

The Automated Patrol Telescope (APT) is a 0.5m telescope of Baker-Nunn design, modified for use with a CCD. It is owned and operated by the University of New South Wales, and located at Siding Spring Observatory, Australia. The current CCD camera images a 2\deg$\times$3\deg\ field with 9.4\arcsec\ pixels. An upgrade is planned (for early 2003) to a pair of 3k$\times$6k CCDs with 4.2\arcsec\ pixels, covering the entire useful field of view of the telescope ($\sim6$\deg\ in diameter). The telescope and dome are computer-controlled, with the possibility of remote or fully automated observation. About 50\% of the observing time on the APT is reserved for this planet search.

\begin{figure}
\plotone{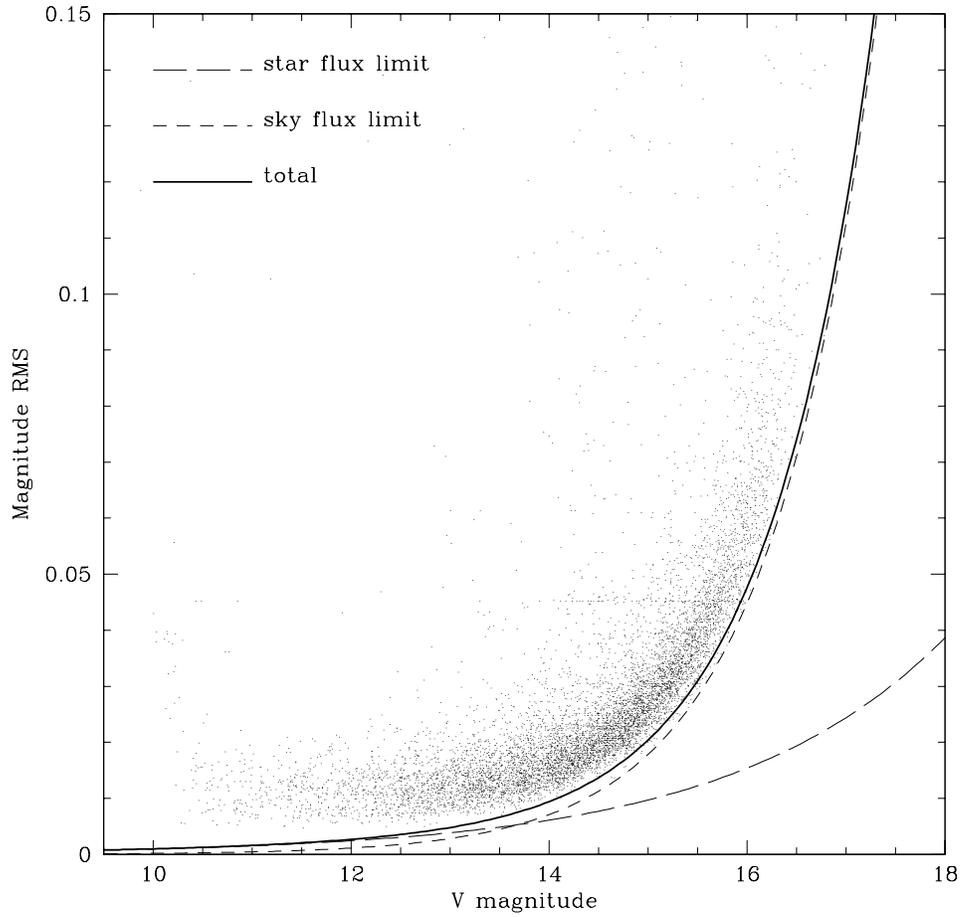}
\caption{Photometric precision for a series of 150 second exposures taken with the APT. The field is centered on the open cluster NGC6633, close to the Galactic plane. The limits due to photon shot noise in the star and sky flux are shown. Readout noise is negligible. Systematic errors limit the precision for bright stars. Stars brighter than $V\sim10.5$ are saturated. }
\end{figure}

A significant drawback of our current system is that the images are undersampled (the FWHM of the point spread-function is 1.3 pixels). Combined with intra-pixel sensitivity variations on the CCD, this causes photometric errors of $\sim3$\% when using standard aperture photometry. Anderson and King (2000) have developed software which deals with the undersampling problem. We are using a modified version of this software. It builds up a model of the ``effective'' point-spread function, which is the instrumental point-spread function convolved with the sensitivity of a single pixel, and measures stars by fitting this model to the image. We can now measure stars down to $V\sim14$ with a relative precision of 1\% (Fig~1) in a single 150 second exposure. For the brightest stars, systematic errors limit the precision. We are working on identifying and eliminating these errors.

We are about to begin regular observations. We select four dense stellar fields near the Galactic plane and cycle between them, taking exposures through a V filter. This way we obtain 5 images of each field per hour, for 6--8 hours per clear night. We observe each field for 2--3 months.

\section{Estimated detection rate}

Kov\'acs, Zucker, \& Mazeh (2002) show that their box-fitting algorithm for detecting transit signals in stellar lightcurves requires an effective signal-to-noise ratio (for the combined measurements during the transit) of 6 for a significant detection. For a typical 1\% deep, 3 hour long transit, we can achieve this by observing just 3 transits. This is true for $\sim1000$ stars with $V<14$ in one low Galactic latitude APT field. To a magnitude limit of $V=14$, near the Galactic plane, F, G, K and M type main sequence stars constitute $\sim60$\% of the stars in the field. Upper main sequence stars are too large for a Jupiter to cause a 1\% transit. Thus, with the 5 times larger field of view of our new CCD system, we will measure $\sim3000$ stars per field with sufficient precision to detect an orbiting Jupiter.

Due to observational constraints (weather, lunar and diurnal cycle, other projects), we estimate that we will detect 10\% of transits which occur in our fields. We are most likely to detect close-orbiting planets with periods of a few days. Almost 1\% of nearby stars host one of these ``Hot Jupiters'' (Butler et al. 2001). Their orbits have a $\sim10$\% probability of being edge-on. We therefore expect to find one planet per $10^4$ stars measured with sufficient precision. We plan to observe 24 fields per year, yielding approximately 7 new planets each year.

\end{document}